\documentclass[amsmath,pra]{revtex4}

\usepackage{graphicx}
\usepackage{dcolumn}
\usepackage{bm}

\begin{document}

\title{\Large Cavity-catalyzed deterministic generation of maximal entanglement
between nonidentical atoms }
\author{\bf Nguyen Ba An}
\email{nbaan@kias.re.kr}
\affiliation{School of Computational Sciences, Korea Institute for Advanced
Study, 207-43 Cheongryangni 2-dong, Dongdaemun-gu, Seoul 130-722, Republic of Korea}

\begin{abstract}
By exactly solving the underlying Sch\"{o}dinger equation we show that one
and the same resonant cavity can be used as a catalyst to maximally entangle
atoms of two nonidentical groups. The generation scheme is realistic and 
advantageous in
the sense that it is deterministic, efficient, scalable and immune from
decoherence effects.
\end{abstract}

\maketitle

\noindent\textbf{1. Introduction}

Quantum entanglement, or simply entanglement for short, though known a long
time ago \cite{qe}, has recently become a key ingredient to perform in
nonclassical ways various tasks of quantum information processing and
quantum computing \cite{r3}. Not only two parties can be entangled with each
other, entanglement exists also between many parties. The two best known
genuine multipartite entangled states are the GHZ-states \cite{ghz} and the
W-states \cite{w} which are inequivalent and exhibit qualitatively different
behaviors (see, e.g. \cite{r7}). Since multipartite entangled states are
necessarily prerequisite resources for quantum network communication and scalable quantum
computers (see, e.g. \cite{r4}) their generation plays a crucial role. A
good deal of generation schemes by diverse mechanisms (see, e.g. \cite
{schemes}) have been studied in detail so far. Here we are concerned with a
cavity-based scheme to generate $N$-partite maximally entangled W-states of
the form 
\begin{equation}
\left| W_{N}\right\rangle =\frac{1}{\sqrt{N}}\left( \left|
e_{1}g_{2}...g_{N}\right\rangle +\left| g_{1}e_{2}...g_{N}\right\rangle
+...+\left| g_{1}g_{2}...e_{N}\right\rangle \right)   \label{WN}
\end{equation}
where $\left| g_{j}\right\rangle $ and $\left| e_{j}\right\rangle $ are the
ground and the excited state of the $j^{th}$ two-level atom. In dealing with
atom-cavity interaction most theoretical schemes have assumed, for
simplicity, identical atoms in the sense that each atom is coupled equally
with the cavity \cite{iden1,iden2}. To achieve error-free results,
synchronization is required that all the atoms be interacted simultaneously
with the cavity: the atoms should be sent at the same time through different
paths (i.e. they enter the cavity at different entrance points). As the
cavity mode has a spatial profile \cite{duan} different atoms actually
experience different couplings with the cavity. In this sense the
experimented atoms are not invariant under atom exchange, or the same, they
are nonidentical with respect to the coupling with the cavity. 
Just very recently 
such practical scenarios have been taken into account for $N=2$ nonidentical
atoms \cite{M1} as well as for an arbitrary $N$ atoms among which $M=1$ atom
is nonidentical with all the remaining $N-1$ identical atoms \cite{M}. This
work carefully considers more general cases of partition of nonidentical
atoms. In section 2 we explicitly prove that neither totally nonidentical
atoms nor totally identical atoms can be maximally entangled by means of
interaction with a resonant cavity. The fully symmetric multi-atom
entanglement of the form (\ref{WN}) can however be generated if the atomic
ensemble possesses a partial asymmetry. Section 3 deals with a bipartite
partition when the atomic ensemble consists of two groups such that the
coupling with the cavity is equal for atoms in the same group but unequal
for atoms in different groups. The generation time is assessed in section 4.
Section 5 is the conclusion.

\vskip 0.5cm

\noindent \textbf{2. Totally nonidentical atoms}

Consider $N$ two-level atoms interacting resonantly with a single-mode
cavity field. In the interaction picture and rotating-wave approximation the
Hamiltonian of the atom-field system reads 
\begin{equation}
H=\sum_{j=1}^{N}f_{j}\left( a^{+}S_{j}^{-}+S_{j}^{+}a\right)  \label{H}
\end{equation}
where $a$ $(a^{+})$ denotes the annihilation (creation) operator of the
resonant single-mode field in the cavity, $S_{j}^{-}=\left|
g_{j}\right\rangle \left\langle e_{j}\right| ,$ $S_{j}^{+}=\left|
e_{j}\right\rangle \left\langle g_{j}\right| $ and $f_{j}$ measures the
strength of the interaction between the $j^{th}$ atom and the same field
mode. The atoms are assumed totally nonidentical in the sense that $%
f_{j}\neq f_{j^{\prime }}$ for $j\neq j^{\prime }.$ The system dynamics
governed by the Hamiltonian (\ref{H}) conserves the so-called excitation
number $\mathcal{N}$ defined as the number of photons plus the number of
excited atoms, i.e. $\mathcal{N}=a^{+}a+\sum_{j=1}^{N}\left|
e_{j}\right\rangle \left\langle e_{j}\right| $. We shall be interested in
the subspace having $\mathcal{N}=1.$ In this subspace there are $N+1$ basic
states which can be chosen as $\left| e_{1}g_{2}...g_{N}\right\rangle \left|
0\right\rangle ,$ $\left| g_{1}e_{2}...g_{N}\right\rangle \left|
0\right\rangle ,$ $...,$ $\left| g_{1}g_{2}...e_{N}\right\rangle \left|
0\right\rangle $ and $\left| g_{1}g_{2}...g_{N}\right\rangle \left|
1\right\rangle ,$ where the latter ket denotes the photon number state. At
any time $t$ the combined atom-field state $\left| \Psi (t)\right\rangle $
can be represented as a linear superposition of the $N+1$ basic states as 
\begin{eqnarray*}
\left| \Psi (t)\right\rangle &=&C_{1}(t)\left|
e_{1}g_{2}...g_{N}\right\rangle \left| 0\right\rangle +C_{2}(t)\left|
g_{1}e_{2}...g_{N}\right\rangle \left| 0\right\rangle \\
&&+...+C_{N}(t)\left| g_{1}g_{2}...e_{N}\right\rangle \left| 0\right\rangle
+C_{N+1}\left| g_{1}g_{2}...g_{N}\right\rangle \left| 1\right\rangle .
\end{eqnarray*}
The differential equations for the coefficients $C^{\prime }s(t)$ can be
derived from the Schr\"{o}dinger equation $i\left| \stackrel{.}{\Psi }%
(t)\right\rangle =H\left| \Psi (t)\right\rangle .$ As a result, we get 
\begin{equation}
\left. 
\begin{tabular}{ccc}
$i\stackrel{.}{C}_{1}(t)$ & $=$ & $f_{1}C_{N+1}(t),$ \\ 
$i\stackrel{.}{C}_{2}(t)$ & $=$ & $f_{2}C_{N+1}(t),$ \\ 
$\cdots $ & $=$ & $\ldots $ \\ 
$i\stackrel{.}{C}_{N}(t)$ & $=$ & $f_{N}C_{N+1}(t),$ \\ 
$i\stackrel{.}{C}_{N+1}(t)$ & $=$ & $\sum_{j=1}^{N}f_{j}C_{j}(t)$%
\end{tabular}
\right\} .  \label{e1}
\end{equation}
The general solution of Eqs. (\ref{e1}) is 
\begin{eqnarray}
C_{k}(t) &=&\frac{f_{k}^{2}\cos (\Omega t)+\sum_{j=1,j\neq k}^{N}f_{j}^{2}}{%
\Omega ^{2}}C_{k}(0)  \nonumber \\
&&+\frac{f_{k}[\cos (\Omega t)-1]}{\Omega ^{2}}\sum_{j=1,j\neq
k}^{N}f_{j}C_{j}(0)  \nonumber \\
&&-\frac{if_{k}\sin (\Omega t)}{\Omega }C_{N+1}(0)  \label{Ck}
\end{eqnarray}
for $k=1,2,...,N$ and 
\begin{equation}
C_{N+1}(t)=-\frac{i\sin (\Omega t)}{\Omega }\sum_{j=1}^{N}f_{j}C_{j}(0)+\cos
(\Omega t)C_{N+1}(0).  \label{CNc1}
\end{equation}
In Eqs. (\ref{Ck}) and (\ref{CNc1}) $\Omega =\sqrt{\sum_{j=1}^{N}f_{j}^{2}}.$
At time $t=t^{\prime }=\pi /\Omega ,$ Eqs. (\ref{Ck}) and (\ref{CNc1})
reduce to 
\begin{equation}
C_{k}(t^{\prime })=\frac{\Omega ^{2}-2f_{k}^{2}}{\Omega ^{2}}C_{k}(0)-\frac{%
2f_{k}}{\Omega ^{2}}\sum_{j=1,j\neq k}^{N}f_{j}C_{j}(0),  \label{Ckt'}
\end{equation}
\begin{equation}
C_{N+1}(t^{\prime })=-C_{N+1}(0).  \label{CNc1t'}
\end{equation}
If the initial system is prepared in a state in which there are no photons
and only one atom, say, the $q^{th}$ atom with $q\in [1,N],$ is excited,
then Eqs. (\ref{Ckt'}) and (\ref{CNc1t'}) simplify to 
\[
C_{q}(t^{\prime })=1-\frac{2f_{q}^{2}}{\Omega ^{2}}, 
\]
\[
C_{p\neq q}(t^{\prime })=-\frac{2f_{p}f_{q}}{\Omega ^{2}}, 
\]
\[
C_{N+1}(t^{\prime })=0. 
\]
Since $C_{N+1}(t^{\prime })=0$ the system at time $t=t^{\prime }$ becomes
decoupled: the cavity returns back to its initial vacuum state but the $N$
atoms turns out to be in the entangled state 
\begin{equation}
\left| \mathcal{A}(t^{\prime })\right\rangle =\left( 1-\frac{2f_{q}^{2}}{%
\Omega ^{2}}\right) \left| ...e_{q}...\right\rangle -\frac{2f_{q}}{\Omega
^{2}}\sum_{p=1,p\neq q}^{N}f_{p}\left| ...e_{p}...\right\rangle  \label{At}
\end{equation}
where $\left| ...e_{j}...\right\rangle $ denotes a state in which only the $%
j^{th}$ atom is excited while all the other $N-1$ atoms remain unexcited.
From Eq. (\ref{At}) it is transparent that the maximally entangled state (%
\ref{WN}) cannot be produced since the atoms are totally nonidentical: the
weight coefficients of $\left| ...e_{j}...\right\rangle $ are all different
from each other. It is worth noting also that even in the case of totally
identical atoms, i.e. $f_{j}=f$ $\forall j,$ maximal entanglement does not
arise because in this case the atoms would appear in the state 
\[
\left| \mathcal{A}^{\prime }(t^{\prime })\right\rangle =\left( 1-\frac{2}{N}%
\right) \left| ...e_{q}...\right\rangle -\frac{2}{N}\sum_{p=1,p\neq
q}^{N}\left| ...e_{p}...\right\rangle 
\]
which is clearly entangled but not maximally.

\vskip 0.5cm

\noindent \textbf{3. Partially nonidentical atoms}

In this section we shall show that the fully symmetric $N$-atom W-state (\ref
{WN}) can be generated if the atomic ensemble possesses a partial asymmetry.
We suppose that the $N$ atoms consist of two nonidentical groups. In group 1
there are $M_{1}$ $(1\leq M_{1}<N)$ identical atoms whereas the number of
identical atoms in group 2 is $M_{2}=N-M_{1}.$ The asymmetry arises from the
fact that each atom of group 1 interacts equally with the cavity mode with a
coupling constant $f_{1}$ while each atom of group 2 also interacts equally
with the cavity mode but with a different coupling constant $f_{2}\neq
f_{1}. $ Under such an asymmetric situation the differential equations for
the coefficients $C^{\prime }s(t)$ read

\begin{equation}
\left. 
\begin{tabular}{ccc}
$i\stackrel{.}{C}_{m}(t)$ & $=$ & $f_{1}C_{N+1}(t),$ \\ 
$i\stackrel{.}{C}_{n}(t)$ & $=$ & $f_{2}C_{N+1}(t),$ \\ 
$i\stackrel{.}{C}_{N+1}(t)$ & $=$ & $M_{1}f_{1}C_{m}(t)+M_{2}f_{2}C_{n}(t)$%
\end{tabular}
\right\}   \label{e2}
\end{equation}
where $m=1,2,...,M_{1}$ and $n=M_{1}+1,M_{1}+2,...,N.$ The general solution
of Eqs. (\ref{e2}) is 
\begin{eqnarray}
C_{m}(t) &=&\frac{M_{1}f_{1}^{2}\cos (\omega t)+M_{2}f_{2}^{2}}{\omega ^{2}}%
C_{m}(0)  \nonumber \\
&&+\frac{M_{2}f_{1}f_{2}[\cos (\omega t)-1]}{\omega ^{2}}C_{n}(0)  \nonumber
\\
&&-\frac{if_{1}\sin (\omega t)}{\omega }C_{N+1}(0),  \label{bm}
\end{eqnarray}
\begin{eqnarray}
C_{n}(t) &=&\frac{M_{1}f_{1}f_{2}[\cos (\omega t)-1]}{\omega ^{2}}C_{m}(0) 
\nonumber \\
&&+\frac{M_{1}f_{1}^{2}+M_{2}f_{2}^{2}\cos (\omega t)}{\omega ^{2}}C_{n}(0) 
\nonumber \\
&&-\frac{if_{2}\sin (\omega t)}{\omega }C_{N+1}(0),  \label{bn}
\end{eqnarray}
\begin{eqnarray}
C_{N+1}(t) &=&-\frac{iM_{1}f_{1}\sin (\omega t)}{\omega }C_{m}(0)  \nonumber
\\
&&-\frac{iM_{2}f_{2}\sin (\omega t)}{\omega }C_{n}(0)  \nonumber \\
&&+\cos (\omega t)C_{N+1}(0)  \label{b}
\end{eqnarray}
where $\omega =\sqrt{M_{1}f_{1}^{2}+M_{2}f_{2}^{2}}.$ At time $t=\theta =\pi
/\omega $ Eqs. (\ref{bm}), (\ref{bn}) and (\ref{b}) reduce to 
\begin{equation}
C_{m}(\theta )=\frac{M_{2}f_{2}^{2}-M_{1}f_{1}^{2}}{\omega ^{2}}C_{m}(0)-%
\frac{2M_{2}f_{1}f_{2}}{\omega ^{2}}C_{n}(0),  \label{cm}
\end{equation}
\begin{equation}
C_{n}(\theta )=-\frac{2M_{1}f_{1}f_{2}}{\omega ^{2}}C_{m}(0)+\frac{%
M_{1}f_{1}^{2}-M_{2}f_{2}^{2}}{\omega ^{2}}C_{n}(0),  \label{cn}
\end{equation}
\begin{equation}
C_{N+1}(\theta)=-C_{N+1}(0).  \label{c}
\end{equation}
If the initial system were prepared in the state 
\begin{eqnarray}
\left| W_{M_{1}}^{-}\right\rangle  &=&-\frac{1}{\sqrt{M_{1}}}\left( \left|
e_{1}g_{2}...g_{M_{1}}g_{M_{1}+1}...g_{N}\right\rangle \right.   \nonumber \\
&&+\left| g_{1}e_{2}...g_{M_{1}}g_{M_{1}+1}...g_{N}\right\rangle   \nonumber
\\
&&\left. +...+\left| g_{1}g_{2}...e_{M_{1}}g_{M_{1}+1}...g_{N}\right\rangle
\right)   \label{WM}
\end{eqnarray}
realized by setting the initial conditions as 
\begin{eqnarray}
C_{1}(0) &=&C_{2}(0)=...=C_{M_{1}}(0)=-\frac{1}{\sqrt{M_{1}}},  \nonumber \\
C_{M_{1}+1}(0) &=&C_{M_{1}+2}(0)=...=C_{N}(0)=C_{N+1}(0)=0,  \label{CM}
\end{eqnarray}
then Eqs. (\ref{cm}), (\ref{cn}) and (\ref{c}) simplify to 
\[
C_{m}(\theta )=\frac{M_{1}f_{1}^{2}-M_{2}f_{2}^{2}}{\omega ^{2}\sqrt{M_{1}}},
\]
\[
C_{n}(\theta )=\frac{2\sqrt{M_{1}}f_{1}f_{2}}{\omega ^{2}},
\]
\[
C_{N+1}(\theta )=0
\]
and the $N$ atoms appear in the entangled state 
\begin{equation}
\left| \mathcal{A}(\theta )\right\rangle =\frac{M_{1}f_{1}^{2}-M_{2}f_{2}^{2}%
}{\omega ^{2}\sqrt{M_{1}}}\sum_{m=1}^{M_{1}}\left| ...e_{m}...\right\rangle +%
\frac{2\sqrt{M_{1}}f_{1}f_{2}}{\omega ^{2}}\sum_{n=M_{1}+1}^{N}\left|
...e_{n}...\right\rangle .  \label{aa}
\end{equation}
If the coupling constants are controlled so that their ratio $r=f_{1}/f_{2}$
satisfies the following constraint 
\begin{equation}
r=1+\sqrt{\frac{N}{M_{1}}}  \label{rct}
\end{equation}
then the state (\ref{aa}) becomes the desired $N$-atom W-state $\left|
W_{N}\right\rangle .$ As recognized from above, the overall process involves
two steps: the first step prepares the initial state (\ref{WM}) and the
second step generates the final state (\ref{WN}). For $M_{1}=1$ the first
step is trivial \cite{M1,M} and the overall process can be looked upon as a
one-step process. Yet, generally, for $M_{1}>1$ preparation of the initial
state (\ref{WM}) is nontrivial. On one hand, we could prepare that state by
using another nonresonant cavity and following the probabilistic scheme
described in \cite{iden2}. More conveniently and more efficiently, on the
other hand, we could use one and the same resonant cavity for doing both of
the steps. Namely, in the first step we send through the empty resonant
cavity $M_{1}$ unexcited atoms of group 1 and one auxiliary atom in the
excited state $-\left| e\right\rangle $ whose coupling constant $f$ 
with the cavity is chosen such that 
\begin{equation}
f=f_{1}\sqrt{M_{1}}.  \label{r}
\end{equation}
Then at time $t=\tau =\pi /f\sqrt{2}$ the cavity turns out to be empty again, the
auxiliary atom jumps down to its ground state but the $M_{1}$ interested
atoms are readily transformed to the required ``initial'' state $\left|
W_{M_{1}}^{-}\right\rangle ,$ Eq. (\ref{WM}). We note that the
just-mentioned first step by using the resonant cavity is measurement-free
and deterministic, in evident contrast with the non-deterministic scheme of 
\cite{iden2} which uses a nonresonant cavity and needs a post-selection
measurement. After the first step we send back the $M_1$ atoms in state 
$\left| W^-_{M_1}\right>$ together with the $M_2$ atoms, all in their 
ground states, to the same cavity to perform the second step. 

\begin{figure}[tbp]
\includegraphics[scale=0.8]{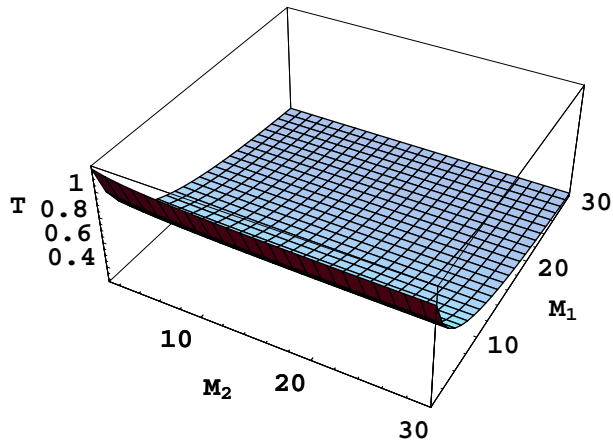}
\caption{Dimensionless total generation time $T=\widetilde{\tau }+\widetilde{\theta }$ 
as a function of $M_{1}$ (the number of atoms in group 1) and 
$M_{2}$ (the number of atoms in group 2 nonidentical with group 1).}
\end{figure}

\vskip 0.5cm

\noindent \textbf{4. Discussion}

To see how long the proposed scheme takes let us scale time in units of $\pi
/f_{1}$ for convenience. That is, we introduce dimensionless time as $%
\widetilde{t}=f_{1}t/\pi $ with $t$ the real time. Then, in terms of $M_{1}$
and $M_{2},$ the (dimensionless) time needed to complete the first step is 
\begin{equation}
\widetilde{\tau }=\frac{f_{1}}{\pi }\tau =\frac{1}{\sqrt{2M_{1}}}
\label{tau}
\end{equation}
whereas the (dimensionless) time needed to complete the second step is 
\begin{equation}
\widetilde{\theta }=\frac{f_{1}}{\pi }\theta =\frac{\sqrt{M_{1}}+\sqrt{%
M_{1}+M_{2}}}{\sqrt{2M_{1}\left( M_{1}+M_{2}+\sqrt{M_{1}(M_{1}+M_{2})}%
\right) }}.  \label{theta}
\end{equation}
Figure 1 plots the dependence of the total (dimensionless) generation time $%
T=\widetilde{\tau }+\widetilde{\theta }$ on $M_{1}$ and $M_{2}.$ The figure
shows that $T$ decreases quickly as $M_{1}$ (the number of atoms in group 1)
increases but slightly as $M_{2}$ (the number of atoms in group 2) increases.

We note that, alternative to the choice (\ref{CM}) of the initial condition,
we could also set 
\begin{eqnarray}
C_{1}(0) &=&C_{2}(0)=...=C_{M_{1}}(0)=C_{N+1}(0)=0,  \nonumber \\
C_{M_{1}+1}(0) &=&C_{M_{1}+2}(0)=...=C_{N}(0)=-1/\sqrt{M_{2}}  \label{CN}
\end{eqnarray}
that correspond to the initial state $\left| W_{M_{2}}^{-}\right\rangle
=-\left( \left| g_{1}...g_{M_{1}}e_{M_{1}+1}g_{M_{1}+2}...g_{N}\right\rangle
+\left| g_{1}...g_{M_{1}}g_{M_{1}+1}e_{M_{1}+2}...g_{N}\right\rangle +...%
\text{ }+\right. $ $\left. \left|
g_{1}...g_{M_{1}}g_{M_{1}+1}g_{M_{1}+2}...e_{N}\right\rangle \right) /\sqrt{%
M_{2}}.$ For the choice (\ref{CN}) we need just making an interchange
between the sub-indices $1$ and $2$ in all the formulae (\ref{rct}) - (\ref
{theta}) as well as in the figure. Comparing the two alternatives suggests
the following strategy to achieve a faster generation time: if $M_{1}>M_{2}$
we choose the conditions (\ref{CM}), if $M_{1}<M_{2}$ we choose the
conditions (\ref{CN}) and if $M_{1}=M_{2}$ either (\ref{CM}) or (\ref{CN})
is equally all-right. Anyway, in a right choice the proposed scheme is robust, taking a
shorter time to entangle a larger number $N$ of atoms. For large enough 
$N$ the total entangling time may become shorter than both the atom
and the photon decay times rendering the scheme immune from decoherence effects.
Since the coupling constant depends on the distance between the central axis
of the cavity and the atom position \cite{duan}, the required ratios of
coupling constants (\ref{rct}) and (\ref{r}) can be experimentally achieved
by appropriately controlling the entrance points through which the atoms
enter the cavity. Furthermore, it is clear that at the beginning the cavity
is set to the vacuum state. At time $t=\tau $ when the ``initial'' state $%
\left| W_{M_{1}}^{-}\right\rangle $ (or $\left| W_{M_{2}}^{-}\right\rangle )$
is prepared the cavity is automatically reset to the vacuum state, i.e. it
is ready for the next step of generating the desired target state $\left|
W_{N}\right\rangle .$ At the end of the second step when the desired state 
$\left| W_{N}\right\rangle $ is generated, the cavity is again reset to its
vacuum state. Hence, one and the same cavity is used for both preparing the
``initial'' state $\left| W_{M_{1,2}}^{-}\right\rangle $ (step 1) and
producing the target state $\left| W_{N}\right\rangle $ (step 2) and, after
each step the cavity is automatically reset to its initial vacuum state,
i.e. the cavity remains unchanged. In this sense, the cavity serves as a
catalyst in the proposed scheme. Finally, our scheme would work similarly
for scalable maximal entanglement of cold ions trapped inside 
a high-Q cavity within
the Lamb-Dicke approximation. In the latter case the ion-cavity couplings
can be controlled by localizing the ions in suitable positions with respect
to the cavity field mode profile.

\vskip 0.5cm

\noindent \textbf{5. Conclusion}

We have considered a realistic situation of atoms interacting with a cavity
with different couplings. We have investigated in detail the case when one
and the same cavity as a catalyst is used to generate $N$-partite W-states
where $N=M_{1}+M_{2}$ with $M_{1}\geq 1$ and $M_{2}\geq 1$ the numbers of
atoms of two nonidentical groups. Our result covers that of $M_{1}=1$ and $%
M_{2}\geq 1$ \cite{M1,M} as a particular case. For $M_{1}>1$ our scheme
proceeds with two successive steps. In the first step an auxiliary atom with
a properly chosen coupling constant is required to prepare the ``initial''
state $\left| W_{M_{1}}^{-}\right\rangle$ or $\left| W_{M_{2}}^{-}\right\rangle,$ 
 depending on $M_1>M_2$ or $M_2>M_1.$ In the second step the desired
target state $\left| W_{N}\right\rangle $ is generated. Both steps are
measurement-free and therefore the whole process, 
as a two-step process, is efficient and
deterministic. The proposed scheme is also scalable with the total
generation time that decreases with increasing $N.$ This feature provides 
an advantage as to overcome the obstacle due to decoherence effects when a large
number of atoms are to be entangled.

\vskip 0.5cm

\textbf{Acknowledgments.} The author thanks KIAS Quantum Information Science
group for support. This research was financed by a Grant (TRQCQ) from the
Ministry of Science and Technology (MOST) of Korea and also by a KIAS R\&D
Fund No 6G014904.

\end{document}